# X-ray diffraction characterization of suspended structures for MEMS applications

P. Goudeau[1], N. Tamura[2], B. Lavelle[3], S. Rigo[4], T. Masri[4], A. Bosseboeuf[5], T. Sarnet[5], J.-A. Petit[4], J.-M. Desmarres[6].

[1] LMP-UMR 6630 CNRS, Université de Poitiers, SP2MI, Bvd Marie et Pierre Curie, BP30179, F-86962 Futuroscope Chasseneuil Cedex.

[2] Lawrence Berkeley National Laboratory, 1 Cyclotron Road, Berkeley 94720, USA

[3] CEMES, UPR 8011 CNRS, 29 rue Jeanne Marvig, F-31055 Toulouse Cedex 4.

[4] ENIT, avenue d'Azereix, BP 1669, F-65000 Tarbes.

[5] IEF, UMR 8622 CNRS, Université Paris XI – Bât. 220, F-91405 Orsay Cedex.

[6] CNES, Agence Spatiale Française, 18 avenue Edouard Belin, F-31401 Toulouse cedex 4.

**ABSTRACT**

Mechanical stress control is becoming one of the major challenges for the future of micro and nanotechnologies. Micro scanning X-ray diffraction is one of the promising techniques that allows stress characterization in such complex structures at sub micron scales. Two types of MEMS structure have been studied: a bilayer cantilever composed of a gold film deposited on poly-silicon and a boron doped silicon bridge. X-ray diffraction results are discussed in view of numerical simulation experiments.

**Keywords:** MEMS, residual stresses, failure, X-ray diffraction, micrometer scale

## INTRODUCTION

Residual stresses can occur in micro devices during process and operation, giving rise to damage and even to failure. These thermo mechanical stresses are difficult to predict from empirical methods and thus experiments and numerical simulation have to be done together in order to improve our understanding of mechanical interaction at meso scales between materials present in these complex systems. Numerous mechanical testing studies have been achieved concerning MEMS structures [1-4] but only a few are dedicated to physical methods such as micro Raman [5], photoluminescence [6] or micro X-ray diffraction [7-8] for probing residual stresses or strains at a micro scale.

In this paper, micro x-ray diffraction technique (µXRD) has been used for studying residual stresses and microstructure at a micron scale in a sub micron thick gold film deposited on a polycrystalline silicon cantilever with micrometer dimensions. Changing white to monochromatic x-ray beam allows grain selection measurement in the gold layer according to their size. The obtained results are compared to macro mechanical measurements and analytical and finite element simulations performed on the same systems. In addition, the technique has been used for studying (100) Si single crystal bridges doped with boron atoms. First results concerning XRD feasibility are given.

**EXPERIMENTAL**

**Gold coated silicon cantilever**: A chip containing test structures such as suspended structures were fabricated according to a complex procedure of micro electronic technologies called CHRONOS [9]. The cantilever is composed of a 0.52 micrometer gold top layer on a polysilicon bottom layer, 1.5 micron thick. The dimensions of the beam are 20 µm large and 100 µm long. Scanning Electron Microscopy (SEM) has been used for characterizing the structures to be analyzed and to identify markers that should be used for calibrating the x-y sample stage (fig. 1).

**Si(B) bridges**: The Boron doped Si bridges have been elaborated at the Institut d'électronique Fondamentale (IEF) at Orsay (France) using laser doping and selective etching [10]. The atomic percentage of boron is high (up to 3-5 %) leading then to strong elastic strains in the silicon unit cell [11]. The beam thickness is of about 200 nm; its width is between 4-6 microns and the length of these micro actuators is comprised between 50 and 500 µm (fig. 2).

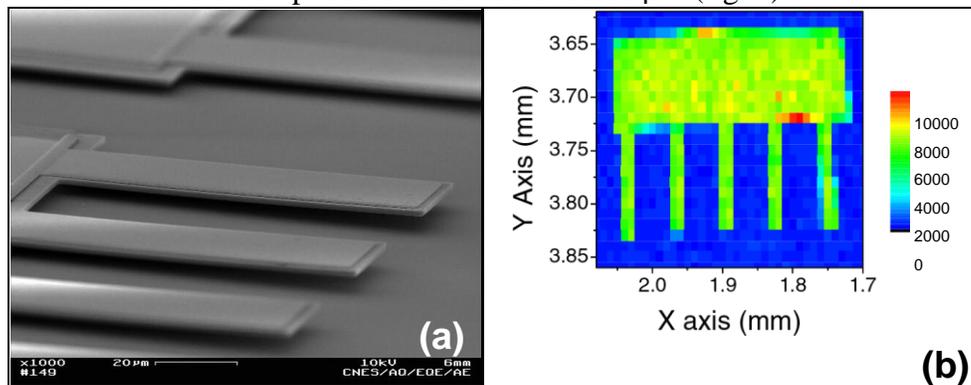

**Figure 1:** Array of bilayer suspended structure test: (a) SEM image and (b) florescence map of the gold top layer.

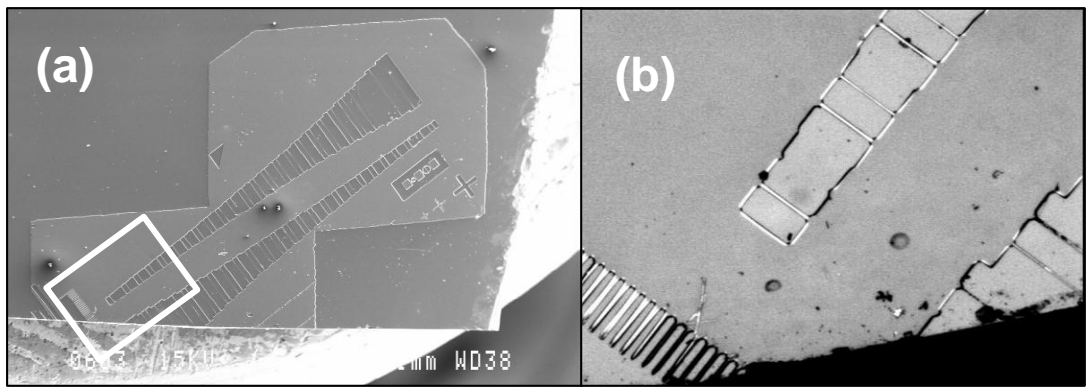

**Figure 2:** SEM images of the Si(B) micro actuators: figure (b) correspond to zoom of the white rectangle in figure (a). The length of the bridges varies between 50 and 500 µm.

**µXRD experiments**: White and monochromatic micro X-ray diffraction measurements have been done at the Advanced light source (ALS) at Berkeley (USA) on the 7.3.3 beam line [10] for measuring stress and grain orientation maps on the micro systems shown on figures 1 and 2. The diffraction patterns are recorded with a 2D CCD detector since rotations of the sample are prohibited. Silicon substrate is used for experiment calibration and the incident x-ray angle is fixed to 45° which means that x-rays entirely penetrate the cantilever and the bridge. Data are analysed using XMAS software developed by N. Tamura for extracting the average stresses.

## RESULTS AND DISCUSSION

**<u>Cantilever</u>**: The suspended structure is shown on fig. 1 (a). The deflection of the free extremity due to residual stresses in the gold film is clearly visible. The array has been then imaged using gold fluorescence on fig. 1 (b). This allows for precisely calibrating the x-y sample stage.

The gold film gives a diffraction signal for both the white and monochromatic x-ray beam(fig. 3). Then, we can consider the coexistence of two grain populations in the gold film volume: large grains (size greater than 0.1 µm) visible with polychromatic x-rays and smaller grains (size down to few nanometers) giving rise to rings (isotropic texture) in the diffraction pattern.

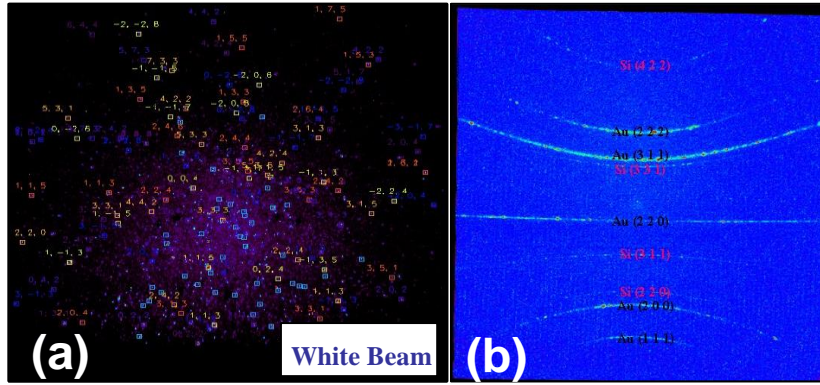

**Figure 3:** Diffraction patterns of gold layer: (a) indexed Laue pattern obtained with white x-ray beam and (b) diffraction rings typical of a diffracting powder. Poly –silicon contribution is only visible in figure (b) which means that all grains are smaller than 0.1 µm.

The first information that can be extracted from white beam measurements is the grain orientation map and then the pole figure. These results are given in fig. 4. We see clearly the presence of a <111> fiber texture with is not really sharp and we also observe fig. 4 (a) a progressive disinclination of the fiber axis when moving from the fixed to the free extremity of the cantilever which can be correlated with the stress induced curvature. The scan is done with a 1 µm$^2$ x-ray beam size and a step of 2 µm.

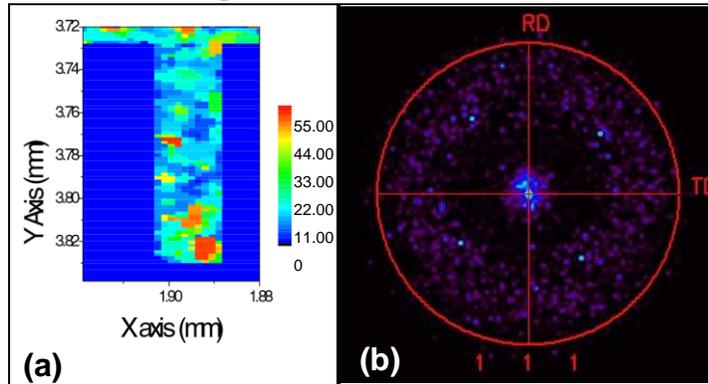

**Figure 4:** Grain orientation measurements in Gold layer: (a) orientation map giving the angular deviation between (111) direction and surface normal and (b) resulting (111) pole figure.

Grain orientation measurements have also been performed with a conventional XRD apparatus using an x-ray beam of 1 mm$^2$ for extracting the three pole figures corresponding to (111), (200) and (220) planes. The results are quite similar indicating a broad distribution of <111> grain fibber axis.

The stress in the region where the cantilever is maintained is approximately equal to 45 MPa (fig 5), a value close to the one determine using a simple mechanical model (Stoney equation) based on the cantilever curvature [9]. The residual stresses are mainly induced by a thermal effect rather than an intrinsic contribution which takes place during the cooling step following the gold film evaporation on the substrate [13]. The polysilicon layer is almost stress free [9]. X-ray diffraction measurements evidence also strong "grain to grain" in plane strain and texture heterogeneities. A similar scan has been done with monochromatic x-ray beam in fig. 6.

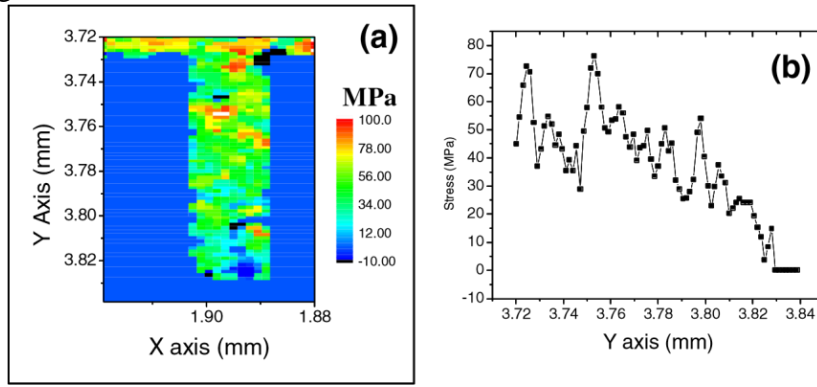

**Figure 5:** Residual stresses in the top gold layer determine in *larger grains* (Φ>0.1µm) with an x-ray beam size of 1µm$^2$: (a) principal in plane stress, (b) mean stress along the cantilever Y axis.

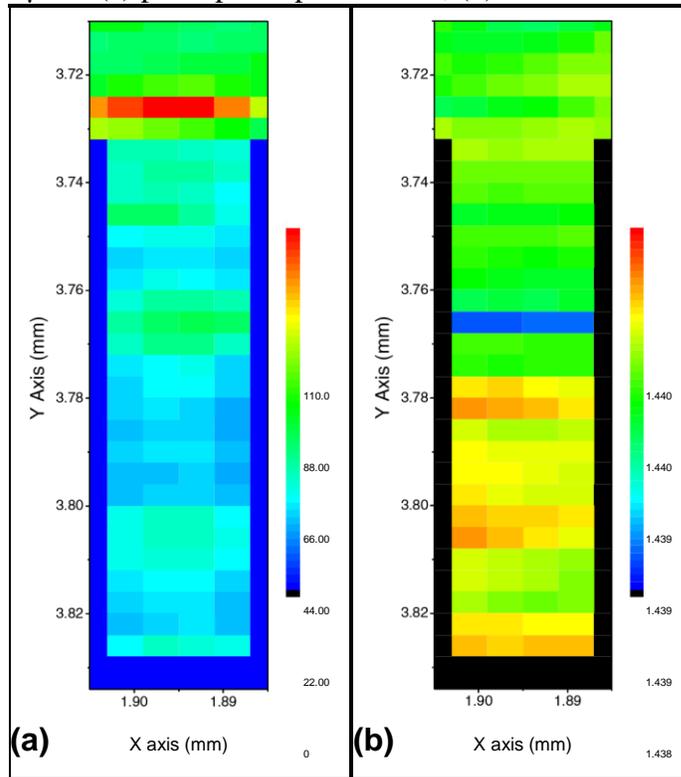

**Figure 6:** Analysis of the (220) diffraction ring of *small gold grains*: (a) integrated intensity (arbitrary units) and (b) d-spacing (Å). The photon energy is about 6 keV and the scan has been realized with a 5 x 3 µm$^2$ spot size on the sample surface and a 4 µm step size.

We observe a slight decrease of the intensity when approaching the free extremity of the cantilever and also a weak increase of the d-spacing. From classical XRD measurements, the reference or unstressed d-spacing value was found to be closed to the bulk gold (1.442 A). The calculated stress is then compressive and increases a bit at the free extremity. This is not surprising since the measured (220) planes in this experiment (45° incident angle and 6 keV photon energy) are parallel to the sample surface. In plane tensile stress produces d-spacing contraction in a direction perpendicular to the sample surface.

Stresses in the cantilever have been simulated numerically by using the Finite Element Code ABAQUS in the approximation of elasticity theory since the deflexion of the beam is small (less than 2 microns) with respect to its length (100 microns). The results are given in fig. 7. The stress in the gold film is almost constant along the cantilever despite strong stresses variations close to the edges. The comparison between simulation and XRD measurements indicates that the apparent stress relaxation in large gold grains at the free extremity of the cantilever is may be not representative of the whole gold film volume. The variation in the smaller grains seems to be opposite but we do not know really the volume fraction of each grain populations.

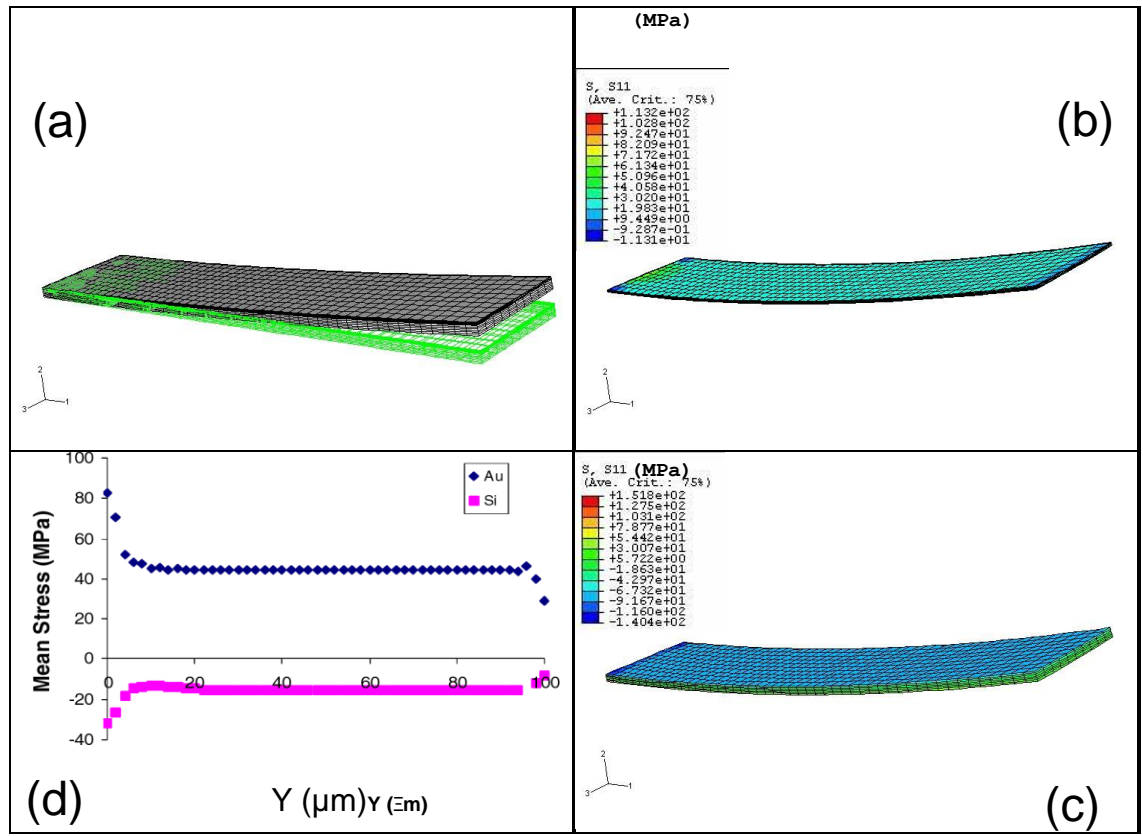

**Figure 7**: Numerical calculation of stresses in the bilayer suspended structure: (a) perspective view of the deformed cantilever (b) and (c) in plane stress map in the gold and silicon layers respectively, (d) stress profiles along the cantilever axis Y in the gold and silicon layers.

**Bridge:** The smallest bridge in fig. 2 (b) has been scanned with a 1 µm$^2$ white beam. The main difficulty here is that the bridge and the substrate structures are identical. However, the two signals may be differentiate on the diffraction pattern shown in fig. 8 (b) owing to the lateral curvature of the bridge which is clearly visible in fig. 8 (a). Observed distortion would result from large strains due to boron atoms which induce Si lattice parameter expansion up to 2.8 %.

## CONCLUDING REMARKS

Micro XRD is a well suited technique for measuring strain/stress in complex architectures encountered in MEMS. Switching white to monochromatic x-ray allows selecting and then analyzing different grain populations of the same phase according to their size. In case of the Si(B) bridge, the Laue signal is shifted because of the (100) plane disorientation with respect to the (100) Si substrate surface. Nevertheless, software developments are needed for extracting quantitative information which will be compared to numerical simulation for improving grain boundary conditions used in modeling.

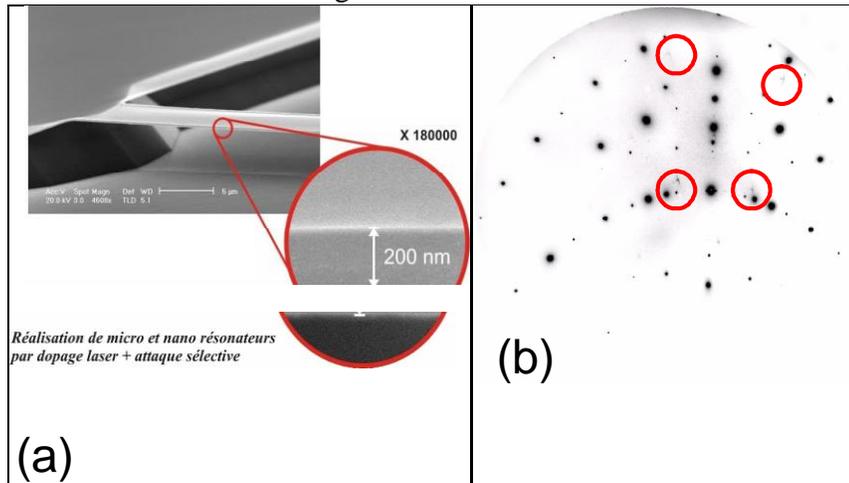

**Figure 8:** Si(B) bridge: (a) SEM image (b) Laue pattern obtained on the bridge. The red circles indicate the position of diffracting intensities related to the bridge.

**Acknowledgements**

The Advanced Light Source is supported by the Director, Office of Science, Office of Basic Energy Sciences, Materials Sciences Division, of the U.S. Department of Energy under Contract No. DE-AC03-76SF00098 at Lawrence Berkeley National Laboratory.